\documentclass[3p]{elsarticle}

\usepackage{hyperref}

\journal{Elsevier}

\usepackage[ansinew]{inputenc}
\usepackage{array}
\usepackage{makecell}
\usepackage{amsthm}
\usepackage{amsfonts}
\usepackage{color}
\usepackage{soul}
\usepackage{rotating}
\usepackage{xcolor,colortbl}
\usepackage{subfiles}
\usepackage{multicol}
\usepackage{stfloats}
\usepackage{url}
\usepackage{ulem}
\definecolor{Gray}{gray}{0.9}
\usepackage{balance}
\usepackage{xcolor}
\usepackage{amsmath}
\usepackage{tikz}
\usetikzlibrary{shapes}
\newcolumntype{C}[1]{>{\centering\let\newline\\\arraybackslash\hspace{0pt}}m{#1}}
\usepackage{rotating}
\usepackage{pdflscape}
\usepackage{afterpage}
\usepackage{capt-of}
\usepackage{longtable}
\usepackage{stfloats}
\usepackage{amsmath,amssymb,amsfonts}
\usepackage{algorithmic}
\usepackage{graphicx}
\usepackage{caption}
\usepackage{textcomp}
\usepackage{longtable}
\usepackage{enumitem}
\usepackage{rotating}
\usepackage{subfigure}
\usepackage{hyperref}
\usepackage{amsxtra}
\usepackage{natbib}
\usepackage{amssymb}
\usepackage{subfigure}
\usepackage{latexsym}
\usepackage{caption}
\usepackage{amsmath,amsfonts,amssymb,amsthm}
\usepackage[ruled,vlined]{algorithm2e}
\usepackage{multirow}
\definecolor{Gray}{gray}{0.9}
\usepackage{balance}
\usepackage{etoolbox}
\usepackage{breqn}

\makeatletter
\def\ps@pprintTitle{%
   \let\@oddhead\@empty
   \let\@evenhead\@empty
   \def\@oddfoot{\reset@font\hfil\thepage\hfil}
   \let\@evenfoot\@oddfoot
}
\makeatother

\begin{document}
\begin{frontmatter}

\title{A Systematic Literature Review of Soft Computing Techniques for Software Maintainability Prediction: State-of-the-Art, Challenges and Future Directions }


\author[1]{Gokul Yenduri}

\author[1]{Thippa~Reddy~Gadekallu*}

\address[1]{School of Information Technology and Engineering, Vellore Institute of Technology, Vellore, Tamilnadu, India\\
(emails: gokul.yenduri@vit.ac.in, thippareddy.g@vit.ac.in)}





\begin{abstract}
The software is changing rapidly with the invention of advanced technologies and methodologies. The ability to rapidly and successfully upgrade software in response to changing business requirements is more vital than ever. For the long-term management of software products, measuring software maintainability is crucial. The use of soft computing techniques for software maintainability prediction has shown immense promise in software maintenance process by providing accurate prediction of software maintainability. To better understand the role of soft computing techniques for software maintainability prediction, we aim to provide a systematic literature review of soft computing techniques for software maintainability prediction. Firstly, we provide a detailed overview of software maintainability. Following this, we explore the fundamentals of software maintainability and the reasons for adopting soft computing methodologies for predicting software maintainability. Later, we examine the soft computing approaches employed in the process of software maintainability prediction. Furthermore, we discuss the difficulties and potential solutions associated with the use of soft computing techniques to predict software maintainability. Finally, we conclude the review with some promising future directions to drive further research innovations and developments in this promising area.

\end{abstract}
\medskip
\end{frontmatter}

\section{Introduction}
\begin{figure}[h!]
    \centering
	\includegraphics[width=.5\textwidth]{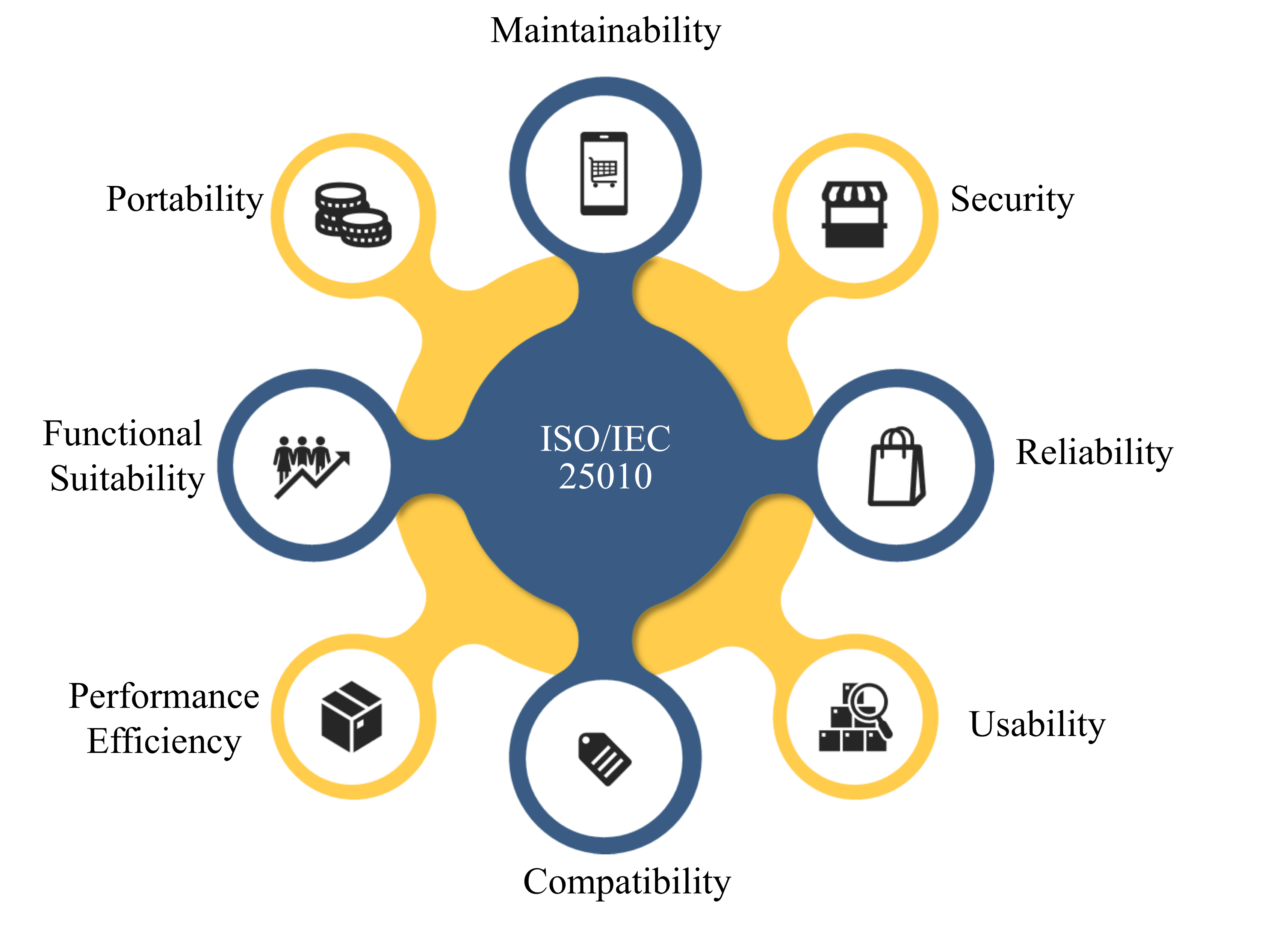}
    \caption{ISO/IEC 25010 Software Quality Model }
    \label{fig:ISO}
\end{figure}

In today's interconnected world, the software is being used in different domains ranging from  financial institutions to power plant management to communication networks to smart cities to mobile applications and many more. Software projects are becoming more challenging to handle due to their dynamic nature \cite{wang2022software}. Even more challenging is the adaptation to the ever-changing requirements of the world. The software must be q ready to adapt solution to an ever-changing environment. Software developers must develop the software by following software development standards  for seamless experience of the end users \cite{noor2022improving}. The software developed should be robust and easy to maintain. The maintainability of the software is essential to the success of a project \cite{butt2022software}. The ISO/IEC 25010 standard specifies eight high-level quality criteria for software, one of which is software maintainability \cite{ISO25010} as depicted in Fig. \ref{fig:ISO}. According to ISO 25000 standards, software maintainability is "the degree of effectiveness and efficiency with which a product or a system can be upgraded to improve it, rectify it, or adapt it to changes in the environment and requirements \cite{iso25000}." Maintainability, as described by the ISO/IEC 25010 quality model, is further subdivided into five subcategories like modularity, reusability, analyzability, modifiability, and testability \cite{peters2020iso}. The fact that software maintainability prediction involves a variety of attributes makes it challenging to quantify effectively. The crucial part of project planning is estimating the resources needed to complete the project, and one way to achieve this is by making predictions. Software maintainability prediction models are trained based on the quality attributes of the historical data \cite{bansal2022cross}. The relationship between internal and external quality attributes is critical to predict software maintainability \cite{elmidaoui2022towards}. Many organizations have realized the role of software maintainability prediction models in better allocation of their resources and budgets \cite{yenduri2021firefly}. Hence, the software maintainability prediction models must be accurate and reliable, which is a challenging task.

\begin{table}[h!]
    \renewcommand{\arraystretch}{1.3}
    \caption{List of key acronyms.} 
    \centering
    \begin{tabular}{|l|l|}
        \hline
        \textbf{Acronyms} & \textbf{Description}\\
        \hline
        {AI}    & {Artificial Intelligence}\\
        \hline
        {CPDP} &{Cross-Project Defect Prediction }\\
        \hline
        {CSA} &{Clonal Selection algorithm}\\
        \hline
        {C\&K} &{Chidamber and Kemerer}\\
        \hline
        {CKJM } &{Chidamber and Kemerer Java Metrics}\\
        \hline
        {FL}    & {Federated Learning}\\
        \hline
        {FA} &{ Firefly Algorithm}\\
        \hline
        {FLANN} &{Functional Link Artificial Neural Network }\\
        \hline
        {GA} &{Genetic Algorithm }\\
        \hline
        {GEP} &{Gene Expression Programming}\\
        \hline
       {ISO}    & {International Organization for Standardization} \\ \hline
        {IEC}    & {International Electrotechnical Commission} \\ \hline
         {LOC} &{Line of Code }\\
        \hline
        {ML}    & 
        {Machine Learning}\\
        \hline
        {MMRE} &{Mean Magnitude of Relative Error}\\
        \hline
         {MAE} &{Mean Absolute Error}\\
        \hline
        {MI} &{Maintainability Index}\\
        \hline
        {OO} &{Object-Oriented}\\
        \hline
        {O\&H} &{Oman and Hagemeister model}\\
        \hline
        {PCA} &{Principal Component Analysis}\\
        \hline
        {PSO} &{Particle Swarm Optimization }\\
        \hline
        {PCC} &{Pearson correlation coefficient }\\
        \hline
          {RSA} &{Rough Set Analysis}\\
        \hline
        {RMSE} &{Random Mean Square Error}\\
        \hline
          {SPMP}    & {Software Product Maintainability Prediction }\\
        \hline
        {SEI} &{Software Engineering Institute}\\
        \hline
        {UIMS} &{User Interface Management System}\\
        \hline
        {QUES} &{Quality Evaluation System}\\
        \hline
         {XAI} &{Explainable Artificial Intelligence}\\
        \hline
    \end{tabular}
    \label{Tab:acronym}
\end{table}

\subsection{Importance of Software Maintainability Prediction}
 Software is dynamic in nature \cite{borg2022agility,manchala2022diversity}. In agile environments, it is essential to prioritize maintainability. Software maintainability is a characteristic that indicates how easily software can be changed and adapted. Software maintainability has become a concern for the software industry as the size and complexity of software systems have increased dramatically \cite{mishra2020devops}.  Maintenance is the most effort-consuming activity in the software life-cycle. It is estimated that over 80\% of the cost spent in the software life-cycle goes into maintenance \cite{papamichail2020generic}. Software maintainability is inversely proportional to maintenance effort. The more effort required to maintain a software system, the lower the value of its maintainability \cite{miloudi2022impact}. A good software maintainability prediction model helps organizations in predicting the maintainability of their software systems with accuracy, enabling them to effectively manage their maintenance resources and drive software maintenance-related decision-making \cite{malhotra2022handling}. As a result, software maintainability prediction models can help in reducing the maintenance effort, cutting the overall time and expense of the software project. If the maintainability of software is assessed early on in the development process, it allows the organizations to provide better resource allocation and planning \cite{jin2022evaluating}. If a software product is difficult to evolve and adapt to new needs, it will not survive in the current market. Any software product developed should be maintainable. If maintainability is not considered throughout the development process prior to releasing the product to the market, it will necessitate large software updates after the product is launched into the market, resulting in enormous losses for the organizations \cite{montano2022countrywide}.\\
After delivering software to the client location, there are a variety of reasons to maintain it as discussed below \cite{uzir2021effects}:
\begin{itemize}
    \item \textbf{Bug Fixing:} It involves searching for errors and correcting them to allow the software to run seamlessly capability enhancement enhancing the software to provide new features required by the customers \cite{winter2022developers}.
    \item \textbf{Replacement:} It involves replacing unwanted functionalities to improve adaptiveness and efficiency \cite{winter2022developers}.
    \item \textbf{Security Issues Fixing:} It involves fixing security vulnerabilities found in your proprietary code or third-party code, especially open source components \cite{canfora2022patchworking}.
\end{itemize}
 Due to the aforementioned reasons, maintainability is very important consideration while building any software product.

\subsection{Contributions and Related Works}
There are very few surveys articles that are totally dedicated towards providing an overview of machine learning (ML) applications to software maintainability prediction as discussed below:

Hadeel Alsolai et al. \cite{alsolai2020systematic} attempted to identify and examine the measurements, metrics, datasets, evaluation measures, individual prediction models, and ensemble prediction models used in the field of software maintainability prediction. They analyzed 56 relevant research papers and 21 conference proceedings. They compared software maintainability to other software quality features, and they concluded that relatively little effort has been put into its prediction. They found out that change maintenance effort and the maintainability index were the most frequently employed software measurements (dependent variables) in the chosen primary studies, whereas class-level product metrics were the most frequently employed independent variables. Also, they stated that several private datasets were utilized in the selected studies, and there is an increasing need for datasets to be made public. They have also found that k-fold cross-validation was used in most regression-related studies and that individual prediction models were used in most research, but ensemble models were used sometimes. They concluded that ensemble models demonstrated increased accuracy in prediction over individual models. They also suggested that there is a need to use ensemble models and other models on a wide range of datasets to make the results more accurate and consistent. However, this work did not provide an extensive assessment of any soft computing  techniques for software maintainability.

Sara Elmidaoui et al. \cite{elmidaoui2020machine} conducted a literature analysis to assess the empirical evidence on the accuracy of software product maintainability prediction (SPMP) using ML approaches. They examined and reviewed the outcomes of 77 selected research papers published between 2000 and 2018 based on maintainability prediction methodologies, validation methods, accuracy criteria, total accuracy of ML techniques, and the top performing techniques. They used the standard procedure for systematic review. The findings suggest that ML approaches are commonly employed for predicting maintainability. In terms of percentage relative error deviation (PRED) and mean magnitude of relative error (MMRE), artificial neural network (ANN), support vector machine/regression (SVM/R), regression and decision trees (DT), and fuzzy and neuro-fuzzy (FNF) approaches are more precise. They stated that the K-fold and leave-one-out cross-validation procedures, as well as the MMRE and PRED accuracy criteria, are widely employed. They also found that ML approaches outperformed non-ML approaches, such as regression analysis (RA) techniques, whereas FNF outperformed SVM/R, DT, and ANN in the majority of studies. They came to the conclusion that, while some ML approaches were thought to be better, none could be called the best. This work did not provide an extensive assessment of any metrics or soft computing techniques for predicting software maintainability.

Ruchika Malhotra et al. \cite{malhotra2020systematic} conducted a systematic review of software maintainability prediction models from January 1990 to October 2019. They evaluated the effectiveness of these models based on several criteria. To achieve the study objective, they have identified 36 research publications. Upon examining these articles, they discovered that different ML, statistical (ST), and hybridization (HB) methodologies were used to construct prediction models for software maintainability. The key conclusion of their study is that the overall performance of ML-based models is superior to that of ST-based models. The application of HB methods to anticipate software maintainability is restricted. This study found that software maintainability prediction models produced using ML approaches performed better than models developed using ST techniques. They came to the conclusion that the prediction performance of a few models built with HB approaches is better, but they couldn't say for sure how well HB techniques work because there hasn't been much research done on them. This work did not provide an extensive assessment of any metrics, datasets, tools, performance measures, or soft computing techniques  for predicting software maintainability.

Sara Elmidaoui et al. \cite{elmidaoui2019empirical} conducted a systematic review of software product maintainability prediction (SPMP) studies in order to examine and synthesise empirical findings about the prediction accuracy of SPMP methodologies in current research. Based on an automated search of nine electronic databases, they conducted a thorough mapping and evaluation of SPMP empirical research published between 2000 and 2018. They categorised 82 main papers of research according to the aforementioned criteria. They found that the most frequently used maintainability predictors were those provided by Chidamber and Kemerer\cite{chidamber1994metrics}, Li and Henry \cite{li1993object}. Their comparative studies revealed that FNF, ANN, MLP, SVM, and group method of data handling (GMDH) techniques provided more accurate predictions than fuzzy and FNF, ANN, and SVM. Based on their findings, the SPMP remains inadequate. They suggested that developing more precise methodologies may enable their use in industry and the attainment of well-formed, generalizable outcomes. In addition, they also offer principles for enhancing the maintainability of software. This work did not provide an extensive assessment of any soft computing techniques for predicting software maintainability.

Justus Bogner et al. \cite{bogner2017automatically} provided a comprehensive review of microservice based system metrics, as conventional metrics, such as object-oriented metrics, are not entirely appropriate in this instance. The chosen metric candidates from the evaluation of the relevant literature were mapped to four major design properties like size, complexity, coupling, and cohesion. Microservice-based Systems (MSBSs) are developed as an agile and selective SBS variant. While the majority of discovered metrics are relevant to this specialisation, the vast number of services, technical variety, and decentralised control have a substantial influence on the automatic collection of metrics in such a system. In order to guarantee the practical application of the proposed metrics to MSBSs, their research suggests that specialist tool for assistance . This work did not provide an extensive assessment of any datasets, tools, performance measures, predication models or soft computing techniques for predicting software maintainability.

Ramon Ablio et al. \cite{abilio2012systematic} discussed the outcomes of a comprehensive literature review conducted to identify contemporary software maintainability metrics for feature-oriented and aspect-oriented technologies. After first screening, the number of detected publications was reduced from 672 to 11. These articles provide 33 and 78 contemporary metrics, respectively, for feature-oriented and aspect-oriented technology. This paper's primary contributions are the list of metrics and measurable properties studies for feature-oriented and aspect-oriented programming, the development of a single library of metrics applicable to both technologies, and the identification of their main references. This work did not provide an extensive assessment of any datasets, tools, performance measures, predication models or soft computing techniques for predicting software maintainability.

Maria Teresa Baldassarre et al. \cite{baldassarre2019software} evaluated empirical studies published over two decades, from 1995 to 2018, in order to answer the question, "What is the most recent empirical evidence about the use of software modelling to support source code maintenance?" To study various facets of the research issue, they conducted a comprehensive literature review of studies published in relevant publications as well as conferences and workshop sessions. In spite of the widespread perception that software models improve the maintainability of source code, this issue has received less attention in the literature, and only a handful of empirical investigations have been undertaken in an industrial context, according to the authors. In addition, the majority of the models utilised are UML. Consequently, their analysis revealed that there is a paucity of research in this sector.

Table \ref{tab:Summary} summarises previous studies. This section highlights the significant contributions of recent work, allowing us to compare them to our own contributions. From the above survey of survey articles, it is evident that there are no studies that focus exclusively on the role of soft computing techniques in predicting software maintainability. To address this gap, this study presents a complete survey of soft computing techniques used in predicting software maintainability. It is greatly anticipated that this study will provide encourage further research from industry and academia.

\begin{table*}[h!]
\centering
\caption{Summary of Literature Survey}
\label{tab:Summary}
\begin{tabular}{|l|l|l|l|l|l|l|l|l|l|l|}
\hline
Ref.& {\rotatebox[origin=c]{90}{~Metrics~}} & {\rotatebox[origin=c]{90}{~Datasets~}} & {\rotatebox[origin=c]{90}{~SPM Models~}} & {\rotatebox[origin=c]{90}{~Tools~}}& {\rotatebox[origin=c]{90}{~ Performance Measures~}}
&{\rotatebox[origin=c]{90}{~Validation ~}}
&{\rotatebox[origin=c]{90}{~ Challenges~}}
&{\rotatebox[origin=c]{90}{~ Possible Solutions~}}
&{\rotatebox[origin=c]{90}{~ Future Directions ~}}\\ \hline
[\cite{alsolai2020systematic}] & \cellcolor[HTML]{d1f8d1}H& \cellcolor[HTML]{d1f8d1}H& \cellcolor[HTML]{FFFFC7}M& \cellcolor[HTML]{d1f8d1}H & \cellcolor[HTML]{d1f8d1}H& \cellcolor[HTML]{FFFFC7}M& \cellcolor[HTML]{FFCCC9}L & \cellcolor[HTML]{FFCCC9}L& \cellcolor[HTML]{FFCCC9}L\\ \hline 
[\cite{elmidaoui2020machine}] & \cellcolor[HTML]{FFCCC9}L & \cellcolor[HTML]{FFFFC7}M& \cellcolor[HTML]{FFFFC7}M& \cellcolor[HTML]{FFFFC7}M & \cellcolor[HTML]{d1f8d1}H & \cellcolor[HTML]{FFCCC9}L & \cellcolor[HTML]{FFCCC9}L& \cellcolor[HTML]{FFCCC9}L & \cellcolor[HTML]{FFCCC9}L\\ \hline 
[\cite{malhotra2020systematic}] & \cellcolor[HTML]{FFFFC7}M& \cellcolor[HTML]{FFFFC7}M&\cellcolor[HTML]{d1f8d1}H & \cellcolor[HTML]{FFFFC7}M&  \cellcolor[HTML]{FFFFC7}M& \cellcolor[HTML]{FFCCC9}L & \cellcolor[HTML]{FFCCC9}L& \cellcolor[HTML]{FFCCC9}L & \cellcolor[HTML]{FFCCC9}L\\ \hline 
[\cite{elmidaoui2019empirical}] & \cellcolor[HTML]{d1f8d1}H& \cellcolor[HTML]{d1f8d1}H& \cellcolor[HTML]{FFFFC7}M& \cellcolor[HTML]{d1f8d1}H & \cellcolor[HTML]{d1f8d1}H& \cellcolor[HTML]{FFFFC7}M & \cellcolor[HTML]{FFCCC9}L& \cellcolor[HTML]{FFCCC9}L & \cellcolor[HTML]{FFCCC9}L\\ \hline 
[\cite{bogner2017automatically}] & \cellcolor[HTML]{d1f8d1}H & \cellcolor[HTML]{FFCCC9}L& \cellcolor[HTML]{FFCCC9}L& \cellcolor[HTML]{FFCCC9}L & \cellcolor[HTML]{FFCCC9}L& \cellcolor[HTML]{FFCCC9}L & \cellcolor[HTML]{FFCCC9}L& \cellcolor[HTML]{FFCCC9}L & \cellcolor[HTML]{FFCCC9}L\\ \hline 
[\cite{abilio2012systematic}] & \cellcolor[HTML]{d1f8d1}H & \cellcolor[HTML]{FFCCC9}L& \cellcolor[HTML]{FFCCC9}L& \cellcolor[HTML]{FFCCC9}L & \cellcolor[HTML]{FFCCC9}L& \cellcolor[HTML]{FFCCC9}L & \cellcolor[HTML]{FFCCC9}L& \cellcolor[HTML]{FFCCC9}L & \cellcolor[HTML]{FFCCC9}L\\ \hline 

[\cite{baldassarre2019software}] & \cellcolor[HTML]{FFCCC9}L& \cellcolor[HTML]{FFCCC9}L& \cellcolor[HTML]{FFCCC9}L& \cellcolor[HTML]{FFCCC9}L & \cellcolor[HTML]{FFCCC9}L& \cellcolor[HTML]{FFCCC9}L & \cellcolor[HTML]{FFCCC9}L& \cellcolor[HTML]{FFCCC9}L & \cellcolor[HTML]{FFCCC9}L\\ \hline 
Our Work & \cellcolor[HTML]{d1f8d1}H& \cellcolor[HTML]{d1f8d1}H & \cellcolor[HTML]{d1f8d1}H & \cellcolor[HTML]{d1f8d1}H  & \cellcolor[HTML]{d1f8d1}H & \cellcolor[HTML]{d1f8d1}H & \cellcolor[HTML]{d1f8d1}H & \cellcolor[HTML]{d1f8d1}H & \cellcolor[HTML]{d1f8d1}H \\ \hline 

\end{tabular}%
\begin{flushleft}
\begin{center}
    
\begin{tikzpicture}

\node (rect) at (0,2) [draw,thick,minimum width=1cm,minimum height=0.7cm, fill= red!25, label=0:Low Coverage] {M};
\node (rect) at (3.4,2) [draw,thick,minimum width=1cm,minimum height=0.7cm, fill= yellow!35, label=0:Medium Coverage] {L};
\node (rect) at (7.5,2) [draw,thick,minimum width=1cm,minimum height=0.7cm, fill= green!20, label=0:High Coverage] {H};
\node (rect) at (11,2) [draw,thick,minimum width=1cm,minimum height=0.7cm, fill= red!75, label=0:Not Applicable] {NA};

\end{tikzpicture}
\end{center}

\end{flushleft}

\end{table*}
\subsection{Systematic Literature Review}
In this work, a systematic literature review approach \cite{xiao2019guidance} is used to understand the use of soft computing techniques for software maintainability prediction that comprises the following steps. First, we discuss the limitations of previously published reviews and the reasons for conducting this review. Next we investigated relevant scientific/research publications on application of several soft computing methodologies in predicting software maintainability. We prioritized peer-reviewed, high-reputation journals, conferences, symposiums, seminars, and books that publish high-quality, peer-reviewed papers from years 1992 to 2022. The references cited in this work are from reputable publications such as IEEE, Elsevier, Wiley, Springer Nature, Taylor, and Francis, as well as reputable research archives and archival websites such as Google scholar and arXiv. In addition, the following search keywords software maintainability, software maintainability prediction, fuzzy logic for software maintainability prediction, fuzzy logic for software maintainability prediction, software change estimation, software product maintainability, evolutionary computing for software maintainability prediction, nature inspired algorithms for software maintainability prediction, soft computing for software maintainability prediction and software maintainability models are employed to extract relevant references and papers concerning soft computing strategies for software maintainability prediction. In the subsequent step, we sort all the retrieved papers by their titles. Low-quality articles were omitted from consideration. Then, we determine the contributions of the papers by reading the abstracts and analyzing the articles for their contributions using the relevant keywords. In the final step, 42 related works are identified for the data necessary for our study of soft computing techniques for software maintainability prediction.

\subsection{Outline of the paper}
\begin{figure*}[h!]
    \centering
	\includegraphics[width=.8\textwidth]{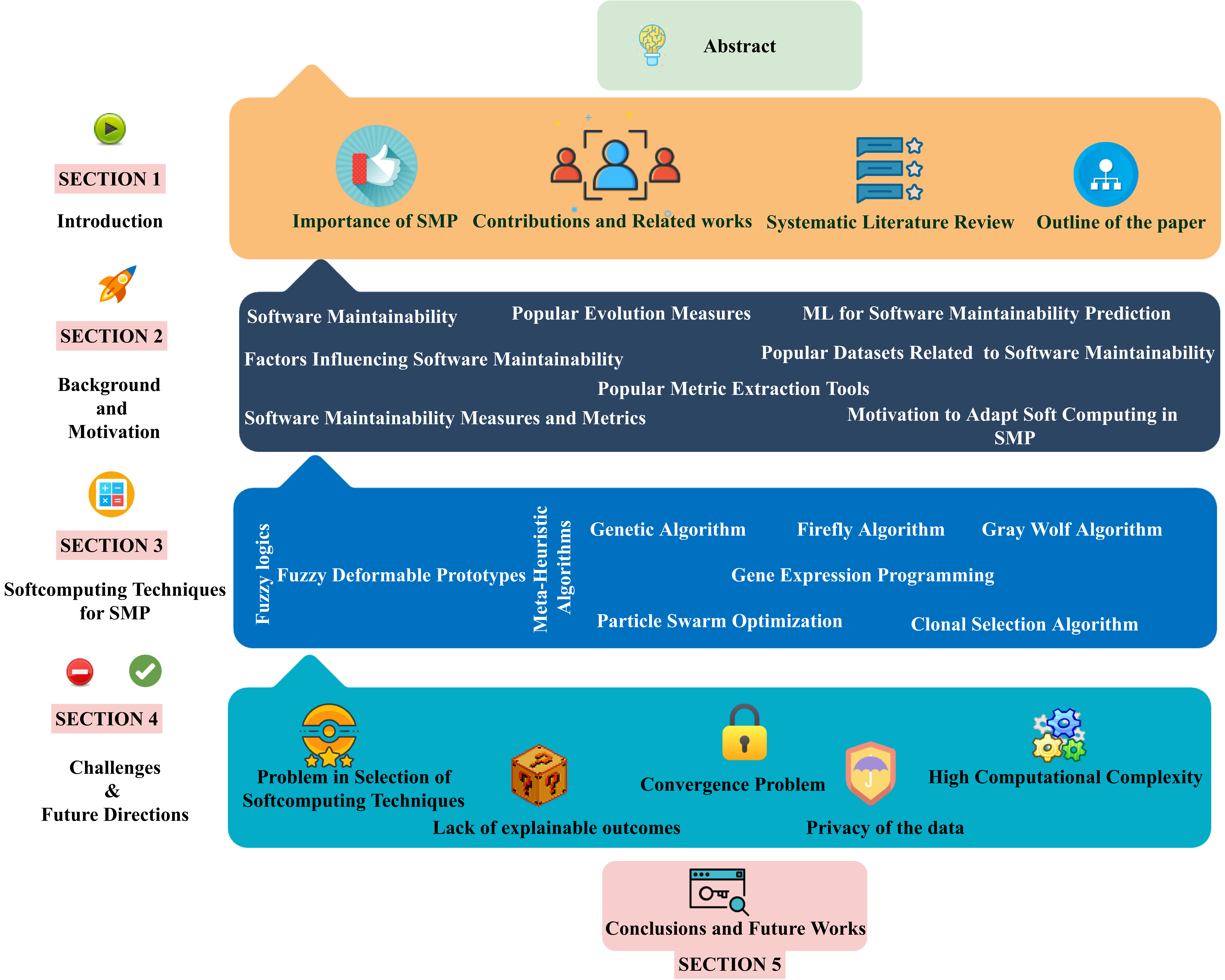}
    \caption{The schematic arrangement of this study}
    \label{fig:flow}
\end{figure*}

The structure of this paper is as follows: Section I is a introductory section about software maintainability. In Section II, we discussed the fundamentals of software maintainability and the motivation for adopting soft computing approaches for the prediction of software maintainability. In Section III, we discussed about the soft computing techniques used in software maintainability prediction. In Section IV, we presented the challenges along with the potential solutions in the use of soft computing techniques for predicting software maintainability. We stated future works and concluded the paper in Section V. The schematic arrangement of this study is depicted in Fig. \ref{fig:flow}, and a list of frequently used acronyms can be found in Table \ref{Tab:acronym}.

\section{Background and Motivation}
In this section, we discuss about influencing factors, measurements, metrics, datasets, evaluation measures, and tools for software maintainability. In addition, motivation to adapt soft computing techniques in the process of software maintainability prediction is also discussed in this section.
\subsection{Software Maintainability}
Software maintenance is an expensive activity that consumes a major portion of the cost of the total project \cite{silva2018technical}. Various activities carried out during maintenance include the addition of new features, deletion of obsolete code, correction of errors, etc. Software maintainability means the ease with which these operations can be carried out \cite{etemadi2022task}. If the maintainability can be measured in early phases of the software development, it helps in better planning and optimum resource utilization \cite{gupta2021optimized}.\\
Below are the various definitions of software maintainability. 

\textbf{Definition 1:} ISO/IEC 25010 defines software maintainability as the degree of effectiveness and efficiency with which a product or system can be modified to improve it, correct it or adapt it to changes in environment, and in requirements \cite{ISO/IEC}.

\textbf{Definition 2:} ISO/IEC 9126 defines software maintainability as a set of attributes that bear on the effort needed to make specified modifications \cite{ISO9126}.

\textbf{Definition 3:} Another standard, ISO/IEC 14764-2006, defines software maintainability as the capability of the software product to be modified, with the objective of software maintenance being to modify existing software while
preserving its integrity \cite{ISO2006}.

\textbf{Definition 4:} Avizienis et.al, defines software maintainability as the ability to undergo repairs and modifications \cite{avizienis2001fundamental}.

\textbf{Definition 5:} Consortium for IT Software Quality (CISQ) defines software maintainability as the degree of effectiveness and efficiency with which a product or system can be modified by the intended maintainers \cite{SM}.

\textbf{Definition 6:} Software Engineering Institute (SEI) defines software maintainability to refer to the property of software that makes these changes possible within acceptable ranges of cost, schedule, and risk \cite{SEI}.

\textbf{Definition 7:} IEEE Standard Glossary of Software Engineering Terminology defines software maintainability as the ease with which a software system or component can be modified to correct faults, improve performance or other attributes, or adapt to a changed environment \cite{159342}. 

\subsection{Factors Influencing Software Maintainability}
The ISO/IEC 25010 quality model provides the basis for the evaluation of software product quality \cite{ISO25010}. The quality model specifies which quality factors are to be used to evaluate the quality of a software product. The quality of a system is the degree to which it satisfies the explicit and implicit needs of its numerous stakeholders \cite{ralph2014dimensions}. The quality model exactly reflects the requirements of various stakeholders (functionality, performance, security, maintainability, etc.) by classifying product quality into factors and sub-factors \cite{motogna2018nlp}. Maintainability is one of those important factors, which is subdivided into five sub-factors as discussed below:
\begin{itemize}
     \item \textbf{Modularity:}
The extent to which a system or computer program is constructed of separate components so that a modification to one component has minimal effect on the remaining components \cite{benkoczi2018design}.

 \item \textbf{Reusability:}
The extent to which an asset may be utilized in one or more systems or for the construction of more assets \cite{saied2018improving}.

\item \textbf{Analysability:}
The effectiveness and efficiency with which it is feasible to analyze the influence of a proposed modification on a product or system, to diagnose a product for defects or failure reasons, or to identify parts to be modified \cite{mikkonen2021machine}.

\item \textbf{Modifiability}
The extent to which a product or system may be modified effectively and efficiently without introducing flaws or compromising product quality \cite{deryugina2019analysis}.

\item \textbf{Testability:}
The extent of effectiveness and efficiency with which test criteria can be set for a system, product, or component, and then with which tests can be carried out to assess whether or not those criteria have been satisfied \cite{garousi2019survey}.

\end{itemize}

\subsection{Popular Software Maintainability Measurement and Metrics}
The most difficult aspect of measuring software maintainability is that it cannot be assessed directly \cite{gradivsnik2020impact}. Consequently, prediction models must rely on indirect measurements. Measuring software maintainability requires a collection of metrics that may be aggregated into a software maintainability function \cite{schnappinger2020defining}. This section describes popular ways to measure software maintainability and metrics that play an important role in predicting software maintainability.
\begin{table*}[h!]
    \renewcommand{\arraystretch}{1}
    \caption{{Li and Henry metrics}}
    \label{tab:LI}
    \centering
    \begin{tabular}{|p{6cm}|p{10cm}|}
        \hline
        \textbf{Metrics} & \textbf{Description}\\
        \hline
        {Weighted Methods per Class (WMC) }  & {It  measure complexity in a class} \\ \hline
        {Depth of inheritance tree (DIT)}    & { It  measures the maximum length between a node and the root node in a class hierarchy} \\ \hline
        {Number of Children(NOC)}  & {It is the number of immediate subclasses of a class} \\ \hline
        {Message Passing Coupling(MPC)}  & {It measures the number of method calls defined in methods of a class to methods in other classes} \\ \hline
        {Response for a class(RFC)}  & {It is the total number of methods that can potentially be executed in response to a message received by an object of a class.} \\ \hline
        {Lack of cohesion in methods(LCOM)} & {It measures the number of not connected method pairs in a class representing independent parts having no cohesion.} \\ \hline
        {Data abstraction coupling(DAC)} & {It measures the coupling complexity caused by Abstract Data Types (ADT)} \\ \hline
        {Weighted methods per class(WMC)}  & { It is the sum of complexities of methods defined in a class} \\ \hline
        {Number of methods(NOM)}  & {It is the total number of methods in a class, including all public, private and protected methods.} \\ \hline
        {Lines of code (SIZE1)}  & {It is total number of lines of code in a class.} \\ \hline
        {Number of properties (SIZE2)}  & { It is total number of attributes and the number of local
methods in a class. } \\ \hline
        
    \end{tabular}
    \label{Tab:Li and Henry metrics}
\end{table*}

\subsubsection{Change Maintenance Effort}

Li and Henry \cite{li1993object} analyzed two projects, UIMS (User Interface Management System) and QUES (Quality Evaluation System), and determined that complexity, coupling, cohesion, and inheritance metrics had a strong correlation with class change volumes. The data regarding maintenance efforts were gathered from two commercial systems with help of Classic-Ada. The data was collected over three years. The maintenance effort is defined by the number of lines per class that has been modified. The metric change may include both additions or removal of lines of code. This proposed measure can be utilized to measure the maintainability of object-oriented systems. The metrics used by Li and Henry are depicted in Table \ref{tab:LI}\\

\begin{dmath}
\mathrm{CHANGE}= \mathrm{F}
(WMC, DIT,
{NOC},{RFC},
LCOM, MPC,
DAC, NOM,\\
SIZE1, SIZE2)  
\end{dmath}

\subsubsection{Maintenance Evaluation by
Maintainability Index}

The Maintainability Index was first introduced in 1992 by Paul Oman and Jack Hagemeister to provide automated software development metrics to help software related decision-making \cite{oman1992metrics}. This method uses a variety of distinct measures to provide a holistic perspective of the relative maintenance effort for various components of a project. The proposed maintainability index is depicted in the below equation. The metrics maintainability index  are depicted in Table \ref{tab:MI}\\

\begin{equation}
\begin{aligned}
MI=171-5.2 \times \ln 
&(\mathrm{HV})-0.23 * \mathrm{CC} \\
&-16.2 \times \ln \mathrm{LOC}+
&50 \times \sin 
&\sqrt{2.4 \times \mathrm{COM}}
\end{aligned}
\end{equation}
\\
The derivative used by SEI \cite{SEIMI} is as follows:

\begin{dmath}
\mathrm{MI}=171-5.2 * \log 2 (\mathrm{~V})-0.23 * \mathrm{G}-16.2 * \log 2 (\mathrm{LOC})+\\ 50 * \sin (\mathrm{sqrt}(2.4 * \mathrm{CM}))
\end{dmath}
The derivative by Microsoft Visual Studio since 2008 \cite{MMI} is as follows:
\begin{dmath}
MI = \operatorname{MAX}\left(0,(171-5.2 * \ln (\text { HV })-0.23 *(\text { CC})-16.2 * \ln (\text { LOC}))^{*} 100 / 171\right)
\end{dmath}

\begin{table*}[h!]
    \renewcommand{\arraystretch}{1}
    \caption{{Maintainability Index metrics}}
    \label{tab:MI}
    \centering
    \begin{tabular}{|p{6cm}|p{10cm}|}
        \hline
        \textbf{Metrics} & \textbf{Description}\\
        \hline
        {Halstead's Volume(HV)} & {It describes the size of the implementation of an algorithm} \\ \hline
        {Cyclomatic Complexity(CC)}  & {It determines the stability and level of confidence in a code} \\ \hline
        {Lines of Code(LOC)}  & { It is a quantitative measurement in computer programming for files that contains code from a computer programming language, in text form.} \\ \hline
        {Percent of Comments(perCOM)}  & {It is a quantitative measurement in computer programming for files that contains comments, in text form.} \\ \hline
        
    \end{tabular}
    
\end{table*}

\subsubsection{Maintenance Evaluation by Change Proneness}
 In software engineering, it is inadequate to rely on a single variable to quantify Object Oriented (OO) metrics; therefore, Chidamber and Kemerer \cite{chidamber1994metrics} define a set of six software metrics (known as the C\&K metric set) that can be used to assess software internal quality using DIT, WMC, NOC, CBO, RFC, and LCOM and prove that there is a linear correlation between them that can influence class change predictions. Maintainability change proneness is TRUE if the change occurs in the class, and FALSE otherwise. The metrics used in maintenance evaluation by change proneness is depicted in Table \ref{Tab:Ck}\\
\begin{equation}
Change Proneness
=\mathrm{F}(\mathrm{WMC},
LCOM, CBO,
DIT, RFC, NOC)
\end{equation}
IF (class change) Change
proneness=
TRUE
ELSE
Change
proneness=
FALSE
\begin{table*}[h!]
    \renewcommand{\arraystretch}{1}
    \caption{{Chidamber and Kemerer metrics}}
    \centering
    \begin{tabular}{|p{6cm}|p{10cm}|}
        \hline
        \textbf{Metrics} & \textbf{Description}\\
        \hline
       {Weighted Methods per Class (WMC) }  & {It  measure complexity in a class} \\ \hline
       {Lack of cohesion in methods(LCOM)} & {It measures the number of not connected method pairs in a class representing independent parts having no cohesion.} \\ \hline
        {Coupling between Objects(CBO)}  & {It is a count of the number of classes that are coupled to a particular class } \\ \hline
        {Depth of inheritance tree (DIT)}    & { It  measures the maximum length between a node and the root node in a class hierarchy} \\ \hline
        {Response for a class(RFC)}  & {It is the total number of methods that can potentially be executed in response to a message received by an object of a class.} \\ \hline
        {Number of Children(NOC)}  & {It is the number of immediate subclasses of a class} \\ \hline
    \end{tabular} 
    \label{Tab:Ck}
\end{table*}

\subsection{Popular Tools for Extracting Maintainability Metrics}
A software metric is a measurement of quantifiable or measurable software characteristics. Software metrics are important for many things, like measuring software performance, planning project tasks, and cost.
There are multiple metrics that are all interrelated for predicting software maintainability. Table \ref{tab:tools} summarized popular specialised tools used by researchers to extract these metrics which are crucial in predicting the maintainability of software.
\begin{table*}[h!]
\centering
\caption{Popular software metric extraction tools}
\label{tab:tools}
\resizebox{\textwidth}{!}{%
\begin{tabular}{|p{4cm}|p{6cm}|p{6cm}|}
\hline
       \textbf{Tools} &  \textbf{Metrics} & \textbf{Description}\\
        \hline
       {Classic-Ada Metric Analyzer \cite{l1993maintenance}}& {DIT, NOC,MPC, LFC, LCOM, DAC, WMC, NOM, SIZE1, SIZE2}&{ $\circ$ A Classic-Ada metric analyzer was developed by Software productivity Solutions, Inc 
        
        $\circ$ It collects
the metrics from the Classic-Ada design and source code} \\ \hline
         {HPMAS prior to perfective maintenance modification \cite{coleman1994using}}& {LOC,  NOM, V(G)}&{ $\circ$ A hierarchical multidimensional assessment model (HPMAS) is
HP's software maintainability assessment
system.

 $\circ$ It is based on a hierarchical organization of a set of software metrics.} \\ \hline
          {SourceMeter static code analysis tool \cite{sourcemeter}}& { More than 50 types of source code metrics \cite{hegedHus2018empirical}}&{ $\circ$ SourceMeter is an innovative tool built for the precise static source code analysis of C/C++, Java, C\#, Python, and RPG projects.
          
           $\circ$ This tool makes it possible to find the weak spots of a system under development from the source code only, without the need of simulating live conditions.} \\ \hline
           {COINS tool \cite{reddy2015software}}& { 13 cohesion metrics, 2
coupling metrics, 19 inheritance metrics and 4 size metrics. In addition, it also gives the values of modifiability,
understandability and testability factors of a class  hierarchy}&{ $\circ$ COINS tool is developed Reddy\cite{reddy2015software}.

 $\circ$ It was created to extract design metric values for hierarchies
in the projects.} \\ \hline
            
               {Chidamber and Kemerer Java Metric (CKJM) tool \cite{spinellis2009ckjm}}& {WMC, DIT, NOC, CBO, RFC, LCOM, Ca, NPM}&{ $\circ$ The program ckjm calculates Chidamber and Kemerer object-oriented metrics by processing the bytecode of compiled Java files.
               
               $\circ$ The program calculates for each class the following six metrics, and displays them on its standard output} \\ \hline
           {Visual Studio \cite{6625879}}& {LOC, HV, CC}&{  $\circ$ It is IDE and helps calculate an index value between 0 and 100 that represents the relative ease of maintaining the code} \\ \hline
            
             {CCCC \cite{CCCC}}& {MI metrics, C\&K metrics, H\&k metrics}&{$\circ$ CCCC is a tool which analyzes C++ and Java files and generates a report on various metrics of the code.}  \\ \hline
             
             {OOMeter \cite{1402128}}& {(RCI, WMC, DIT, NOC, CBO, RFC, LCOM, TCC, LCC, LOC }&{$\circ$ A software metric tool that accepts Java and C\# source code as well as UML models in XMI format.} \\ \hline
             {IntelliJ IDEA \cite{IDE}} & {24 different kinds of metrics}&{ $\circ$ It is IDE with touring plugins it helps to calculate software metrics} \\ \hline

    \end{tabular}%
}
\end{table*}

\subsection{Popular Software Maintainability Datasets}
The popular datasets used for software maintainability prediction in the state-of-the-art are discussed in this section as summarized in Table \ref{tab:dataset}.

\begin{table*}[h!]
\centering
\caption{Popular datasets for software maintainability Prediction}
\label{tab:dataset}
\resizebox{\textwidth}{!}{%
\begin{tabular}{|p{4cm}|p{6cm}|p{6cm}|}
\hline
       \textbf{Dataset} &  \textbf{Dataset Source} & \textbf{Description}\\\hline
{QUES \cite{li1993object}} &  {Commercial software
products (Quality Evaluation
System)} & {  $\circ$ It is first published in the year 1993 by li and henry in there work on object-oriented metrics that predict maintainability.

$\circ$  QUES consists of 71 classes.}\\\hline
{UMIS \cite{li1993object}} &  {Commercial software
products (User Interface
System)} & {  $\circ$ It is first published in the year 1993 by li and henry in there work on object-oriented metrics that predict maintainability.

$\circ$  UMIS consists of 39 classes.}\\\hline

{NASA \cite{NASA1,NASA2}}& {NASA Metrics Data Program} & { $\circ$ These datasets are developed from the metrics related to  NASA spacecraft  }\\\hline

{UML class diagram \cite{Genero}}& {Bank Information Systems} & { $\circ$ It consists of twenty eight UML class diagrams of the same universe of discourse, related to bank information systems }\\\hline

{ Apache lucene \cite{lucene}}& {Apache Lucene (module: Core)} & { $\circ$ The Apache Lucene project develops open-source search software and reqiured metrics should be extracted using tools. \cite{yenduri2021firefly}}\\\hline
           
    \end{tabular}%
}
\end{table*}

\subsection{Evaluation Measures}
A crucial step in any empirical study is determining the accuracy of the predicted model. MRE, MMRE, MAE, RMSE, are the popular evaluation measures used various researchers in process of predicting maintainability of software with help of soft computing techniques.\\

\textbf{Mean Relative Error (MRE):} The relative error is calculated by dividing the absolute error by the estimated value.
\begin{equation}
M R E=\frac{\mid \text { actual } i \text {-estimated } i \mid}{\text { actual } i}
\end{equation}
\textbf{Mean Magnitude of Relative Error (MMRE):} MMRE is the difference between actual and estimated effort relative to the actual effort. The mean takes into account the numerical value of every observation in the data distribution, and is sensitive to individual predictions with large MREs.
\begin{equation}
M M R E=\frac{1}{N} \sum_{\chi=1}^N M R E
\end{equation}
\textbf{Mean Absolute Error (MAE):} MAE is the is a calculated of errors between paired observations expressing the same phenomenon. 
\begin{equation}
\mathrm{MAE}=\frac{\sum_{i=1}^n\left|y_i-x_i\right|}{n}
\end{equation}
\textbf{Root Mean Square Error (RMSE)}
RMSE is the standard deviation of the residuals (prediction errors).
\begin{equation}
R M S E=\sqrt{\frac{\sum_{i=1}^N\left(\text { Predicted }_i-\text { Actual }_i\right)^2}{N}}
\end{equation}

\subsection{Machine Learning for Software Maintainability Prediction }
Software maintainability is one of the key criteria of the software quality model proposed by the Software Engineering Institute (SEI). Numerous studies have shown that software maintainability is one of the most difficult measurements of software quality due to the presence of multiple metrics, not all of which can be used to predict future maintainability \cite{ardito2020tool}. Measurement of software maintainability is based on a collection of metrics that can be aggregated into a software maintainability measure. Measures such as the maintainability index and change proneness can be utilized in predicting the maintainability of software. The use of machine learning for the maintainability prediction of software has gained prominence. ML can predict software maintainability with greater precision without being explicitly programmed. ML uses historical software maintainability data to predict future maintainability.

\subsection{Motivation to Adapt Soft Computing in Software Maintainability Prediction}
Zadeh coined the term "soft computing" in the year 1992 \cite{dinesh2022soft}. Soft computing techniques are inspired by human reasoning and have the capacity to deal with imprecision, uncertainty, and partial truth \cite{zhang2022soft}. The essence of soft computing is its effort to accommodate the pervasive imprecision of the real world, in contrast to traditional methods. Thus, the guiding idea of soft computing is to leverage the tolerance for imprecision, uncertainty, and partial truth \cite{ibrahim2016overview}.  Soft computing techniques help in achieving tractability, robustness, cheap solution cost, and a more accurate reflection of reality in predicting the maintainability of software \cite{jain2016empirical}. Soft computing is not a singular approach, it is a collaboration of several approaches such as fuzzy logic, neurocomputing, genetic computing, probabilistic computing, nature inspired computing, etc \cite{yardimci2009soft}.
Soft computing techniques can help software maintainability in the following ways
\begin{itemize}
     \item \textbf{Software Metric Selection:} Feature selection or metric selection can be done using soft computing techniques. This helps developers to identify appropriate metrics for measuring the maintainability of any software.
     \item \textbf{Improves Prediction Performance of Software Maintainability:} Soft computing techniques can help reducing the training time by the hyper-parameter tuning the data, whereby improving the performance of prediction models.
     \item \textbf{Reduces Computation Cost:}
     The performance of the model is directly proportional to the computational cost of the software. Hence, the computational cost of maintainability reduces as the performance of the ML model increases. 
     
\end{itemize}

\section{Soft Computing Techniques for Software Maintainability Prediction}
Soft computing, in contrast to conventional computing, deals with approximation models and provides answers to difficult real-world issues. Soft computing is tolerant of imprecision, ambiguity, partial truth, and approximations, as opposed to hard computing. In essence, the human mind is the model for soft computing. Soft computing is founded on methods including fuzzy logic, GA, ANN, ML, and expert systems \cite{zhang2022soft}. Soft computing theory has become a prominent topic of study in software maintainability. Soft computing approaches are currently being utilized successfully in a variety of healthcare, industrial, and other applications \cite{dote2001industrial}. It is evident that soft computing techniques and application areas will continue to increase. Soft computing is described as a series of computer approaches based on artificial intelligence (human-like decision making) and natural selection that gives a rapid and cost-effective solution to very complicated problems for which analytical formulations are unavailable. The objective of soft computing is to discover an accurate approximation that yields a resilient, computationally efficient, and cost-effective solution while reducing computational time. The vast majority of these strategies are primarily motivated by biologically inspired occurrences and social behavior patterns. The emergence of soft computing into the technical world was facilitated by research in bio inspired algorithms, fuzzy logic, and GA. Soft computing also includes swarm intelligence, PSO, and a number of other techniques. The softcomputing techniques used for software maintainability prediction is summarized in Table \ref{tab:m2}. Rest of this section discusses about several soft computing techniques that can be used for software maintenance prediction. 

\subsection{Fuzzy Logic}
Fuzzy logic is a technique for solving issues that are too complicated to be quantified. Zadeh introduced the fuzzy set theory in 1965 \cite{charisis2022fiseval}. It gives a strategy for dealing with imprecision and information granularity. It is a mathematical approach to deal uncertainty. The primary benefit of this method is that it is less reliant on historical data. Fuzzy logic models can be created with little or limited data. A fuzzy logic system that accepts imprecise data and ambiguous statements as low, medium, or high and makes judgments based on them. There are four significant modules. The fuzzification module modifies the input values to be fuzzy values. These are then processed in the fuzzy domain by an interface engine based on an expert-supplied knowledge base (rule base). Finally, defuzzification converts fuzzy domain processed data to crisp domain \cite{reddy2018heart,asghar2021senti}.

Amrendra Pratap et al. \cite{pratap2014estimation} used fuzzy logic to estimate the maintainability of software systems. They claim that fuzzy logic-based approaches have several advantages over other techniques such as neural networks. They have used adaptability (AD), complexity (CLX), understandability (USD), documentation quality (DocQ), and readability (RD) as inputs, while maintainability was considered as an output. In total, 243 fuzzy sets have been designed and are represented by membership functions (Low, Meadium, and High) of MATLAB. Output is represented by very easy, easy, medium, high, and very high. The defuzzification process generated the maintainability value corresponding to the input sets. Their results show that the proposed model can be used to predict the maintainability of software.

Kumar et al. \cite{kumar2017software} considered a hybrid of neural network and fuzzy logic to create a maintainability model with ten distinct OO static source code metrics as input. This method is applied to data regarding the maintainability of two commercial software products, such as UIMS and QUES. Rough set analysis (RSA) and principal component analysis (PCA) are utilised to select an appropriate set of metrics from the ten metrics used to enhance the performance of the maintainability prediction model. The results of experiments yielded MMRE of 0.2826 and 0.3375 respectively for UIMS and QUES which indicate that the Neuro-Fuzzy model can accurately predict the maintainability of OO software systems. When the number of computing nodes is increased following the implementation of the concept of parallel computing, the training time is observed to be significantly reduced. Moreover, it was discovered that a subset of metrics selected using feature selection techniques, i.e., PCA and RSA, could accurately predict maintainability.

\subsubsection{Fuzzy Deformable Prototypes}
FDP is defined as a linear combination of adaptable Fuzzy Prototypical Categories (expressed as attribute tables) to any actual circumstance, where the coefficients represent the degrees of belonging to each of these Fuzzy Prototypical Categories \cite{olivas2017some}. Expanding the combination provided by the idea of Deformable Prototype to include affinity with many Fuzzy Prototypical Categories, the description of a true scenario would be:\\

\begin{equation}
C_{\text {real }}\left(w_{1} \ldots w_{n}\right)=\left|\sum \mu p_{i}\left(v_{1} \ldots v_{n}\right)\right|
\end{equation}
\\
where:\\
$C_{\text {real }}  \quad$ = Real case.\\
$\left(w_{1}, \ldots, w_{n}\right)$ = Parameters describing the real case.\\
$p_{i} \quad$ = Degrees of compatibility with Fuzzy Prototypical Categories different to 0 .\\
$\left(v_{1}, \ldots, v_{n}\right)$ = Parameters of these Fuzzy Prototypical Categories.\\

The process of developing object-oriented information systems (OOIS) is becoming more and more complex, which requires continuous monitoring and evaluation of class diagrams. The introduction of UML as a standard for object modeling is a significant development. Although it does not ensure the system's quality, it is an excellent starting point. To evaluate the quality of an OOIS, measurements are needed from the commencement of its development. The FDP have been successfully used for several real-world problems, including forest fire prediction, financial analysis, and medical diagnostics. This motivated Genero et al. \cite{Genero} to adapt fuzzy deformable prototypes to predict the UML class diagram maintainability based on the metrics values and the expert's rating of each of the maintainability sub-characteristics.

In summary, fuzzy logic can be used to determine the relationship between software metrics and software maintainability. Although fuzzy logic can aid in predicting the maintainability of software, it is unable to precisely identify the flaw in the code that affected the software's maintainability. The explainability of the results is difficult to understand due to their black box nature. As the data is stored in a single database, there is also a risk to the privacy of user-shared data.

\subsection{Meta Heuristics}
 A metaheuristic is a higher-level procedure or heuristic that is used to find, create, or choose a heuristic that may provide a good enough solution to an optimization problem, especially when there isn't enough information or there isn't enough computing power. Metaheuristics take a sample of a subset of solutions that would be too many to list. Metaheuristics may not make many assumptions about the optimization problem they are trying to solve, so they may be useful in a wide range of situations. Metaheuristics are different from optimization algorithms and iterative approaches in that they do not guarantee that there is a globally best solution for a certain type of problem \cite{tripathy2022harris}. Many metaheuristics use some kind of stochastic optimization, which means that the solution found depends on the random variables that are made. In combinatorial optimization, metaheuristics can often find good answers with less computing work than optimization algorithms, iterative approaches, or basic heuristics by searching through a wide range of possible solutions \cite{hussain2019metaheuristic, dragoi2021review}. Some of the popular metaheuristic approaches used for software maintainability are discussed in the following subsections.

\subsubsection{Bio-inspired algorithms}
Bio-inspired computing algorithms are based on the principles and inspiration of the biological evolution of nature in order to develop new and robust competing techniques. Genetic algorithm, firefly algorithm and grey wolf algorithm are the examples of bio-inspired computing algorithms \cite{fan2020review}. Some of the popular bio-inspired algorithms used by several researchers for software maintainability prediction are discussed below.

Genetic Algorithms (GA) was once such algorithm introduced by John Holland in 1975 \cite{lee2022different}. A GA is a search heuristic influenced by Charles Darwin's idea of evolution through natural selection. This algorithm resembles the process of natural selection, in which the fittest individuals are chosen for reproduction to generate the next generation. The natural selection process begins with the selection of the healthiest individuals from a population. They generate offspring that inherit their parents' qualities and are added to the following generation. If parents are more fit, their offspring will be better and have a greater chance of survival. This procedure is repeated until a generation containing the fittest individuals is produced. This method is useful for solving search problems \cite{eshelman2018genetic,kandati2022genetic,katoch2021review}.

Lov Kumar et al. \cite{KUMAR2015798} attempted to predict software maintainability using a subset of class-level OO metrics. A Neuro-Genetic (Neuro-GA) approach with a model with 10-fold and 5-fold cross-validation was developed for the QUES and UIMS software. To analyze the software maintainability of QUES and UIMS software, four analysis approaches employing different metric sets were evaluated. The software metrics serve as input data for training the network and predicting the maintainability of the software product. Compared to the previous work \cite{zhou2007predicsting, van2006application}, this study proves that the Neuro-GA approach produced promising outcomes with MMRE of 0.3155 and 0.3775 respectively. In addition, the results revealed that the Neuro-GA approach selected subset of metrics improved the accuracy of maintainability prediction. In another effort to use software metrics for constructing object-oriented software maintainability prediction models, the researchers employed software metrics. To build models, AI hybrid techniques, such as the hybrid approach of functional link artificial neural network (FLANN) and GA, were considered. QUES and UIMS software were also subjected to 10-fold and 5-fold cross-validation approaches. The results showed that in the case of UIMS, the proposed model FLANN-Genetic Approach (FGA) by taking into account a smaller set of object-oriented metrics of RST as input, provided better results compared to the other methodologies in this proposed work \cite{kumar2016hybrid}. Ashu Jain et al. \cite{jain2016empirical} explored two versions of four open source software and computed the change for each. Using the ckjm tool, the independent metrics were computed. With the help of ML classifiers provided by the Weka tool, the prediction models were constructed. The outputs of the suggested classifiers were then combined and examined using MAE and RMSE. It was concluded from their findings that the GA with RMSE of 0.045, 0.1091, 0.0128, and 0.0393 for Jtds, Jwebunit,Jxls, and Sound helix open source software perdicts better than the other ML classifiers used in their research. In 2019, Lov Kumar et al. in \cite{kumar2019estimation} presented a technique for predicting the maintainability of OO software using three distinct neural networks and 10 class-level metrics. The GA with the gradient descent algorithm is used to train the GA-GD hybrid NN model, which outperformed other algorithms considered in their study.

In summary, the GA is an effective soft computing method for predicting the maintainability of software. GA requires enormous computational capacity for massive datasets, which is challenging. Additionally, the difficulty of the convergence problem in GA is significant. The GA is sensitive to initial conditions and can converge to local optimal solutions on occasion. A software maintainability prediction with the help of GA is difficult to understand and interpret, which makes it challenging to comprehend why a particular solution is attained.

The firefly algorithm (FA) was developed by Xin-She Yang in 2008 \cite{yang2020firefly}. The FA imitates the social behavior of summertime fireflies. Specific characteristics of fireflies include communication, prey searching, and mating. Fireflies are unisexual, fireflies are attracted to each other regardless of gender. Attractiveness is directly proportional to brightness, the lower-brightness firefly approaches the higher-brightness firefly. If no other firefly is brighter than the present one, it will go aimlessly throughout space. The relationship between brightness and the cost function is significant. In maximizing problems, the brightness is proportional to the value of the cost function.

In 2021, Gokul et al. \cite{yenduri2021firefly} proposed firefly based maintainability index by taking into account certain software metrics with the aim to minimize error. The FA is compared to other traditional models such as DE, Artificial Bee Colony, PSO, and GA in terms of performance metrics such as differential ratio, correlation coefficient, and RMSE. The proposed model has shown promising outcomes. The proposed firefly method outperformed both existing conventional and meta-heuristic models.

In summary, FA is more effective at handling multi-modal optimization problems and nonlinear problems. Additionally, it does not employ velocity. Therefore, there is no issue with velocity variation. Concurrently, FA has a rapid convergence rate for locating global optimization. In addition, similar to other GA, FA does not require the optimal initial solution to initiate the iteration phase. In addition, it may be easily combined with other optimization approaches to build hybrid algorithms. It has issues related to high computational complexity, convergence speed, justification of results and privacy of the data shared.

Grey wolf optimization algorithm (GWO) is new meta-heuristic optimization technology proposed by Mirjalili et al. \cite{mirjalili2014grey}. Its principle is to imitate the behavior of grey wolves in nature to hunt in a cooperative way. This motivated Gokul et al. to allot an optimal weight and a constant to each software metric, which is optimized by grey wolf optimization (GWO). As a result, it can provide a new variant of MI by proposed enhanced model-GWO (EM-GWO) \cite{yenduri2021nonlinear}. This optimized MI can ensure the efficiency of the respective software in such a way that it can provided an enhanced maintainability measure of the software. Further, the proposed method is compared with conventional models such as GA, PSO, ABC,DE, and FF, and the results proved EM-GWO has produced better results.

In summary, Although EM-GWO  is better for a small dataset but it has a disadvantages of convergence speed, low solution accuracy, and easy to fall into the local optimum.

\subsubsection{Particle Swarm Optimization}
Particle swarm optimization (PSO) is based on the method given by Kennedy and Eberhart \cite{shami2022particle}. The particle swarm algorithm begins by constructing and assigning starting velocities to the first particles. It calculates the optimal (lowest) function value and particle placement by evaluating the objective function at each particle location. It selects new velocities depending on the existing velocity, the optimal positions of the particles, and the optimal positions of their neighbors. It then iteratively updates the particle positions, velocities, and neighbors (the new location is the previous one plus the velocity, tweaked to keep particles inside boundaries). The algorithm iterates until it meets a stopping condition. 

In 2016, Lov Kumar et al. in \cite{kumar2016hybrid} used software metrics for constructing object-oriented software maintainability prediction models. To build models, AI hybrid techniques, such as the hybrid approach of FLANN and Particle swarm optimization, were considered. QUES and UIMS software were also subjected to 10-fold and 5-fold cross-validation approaches. The results showed that in the case of QUES, the proposed model FLANN-PSO (FPA) by taking into account a smaller set of object-oriented metrics of RST as input, provided better results compared to the other methodologies in this proposed work. 

In summary, PSO performs better with smaller maintainability datasets. Larger datasets present Challenges for the PSO because of its slow convergence speed and large search space. The issue related to the privacy of data and the interpretation of the software maintainability prediction of the software is another Challenge in adapting the PSO.

\begin{table*}[h!]
\centering
\caption{Soft Computing Techniques for Software Maintainability Prediction}
\label{tab:m2}
\resizebox{\textwidth}{!}{%
\begin{tabular}{|l|p{3cm}|p{2cm}|p{2cm}|p{2cm}|p{3cm}|p{2cm}|p{2cm}|}
\hline
        \textbf{Ref}& \textbf{soft computing Approach Used}& \textbf{Performance measure Used} & \textbf{Tool Used}&\textbf{Dataset Used}&\textbf{Type of maintenance Used}&\textbf{Validation}\\
        \hline
        {\cite{olivas2017some}}& {FDP Approach}& {Mean}&{Survey questionnaire}&{Bank Information Systems}&{UML metric model}&{K-fold} \\ \hline
        
        {\cite{kumar2017software}}& {Neuro-fuzzy Approach}& {MMRE}&{Classic-Ada Metric Analyzer}&{QUES and UIMS}&{Change maintenance effort}&{K-fold} \\ \hline
        
        {\cite{KUMAR2015798}}& {Neuro-GA Approach}& {MMRE, MAE, RMSE, SEM}&{Classic-Ada Metric Analyzer}&{QUES and UIMS}&{Change maintenance effort}&{K-fold} \\ \hline
        
         {\cite{kumar2016hybrid}}& {FLANN-GA Approach}& {MMRE, MAE}&{Classic-Ada Metric Analyzer}&{QUES and UIMS}&{Change maintenance effort}&{K-fold}  \\ \hline
         
         {\cite{kumar2019estimation}} & {Neuro-Fuzzy Approach}& {MMRE, MAE, SEM }&{Classic-Ada Metric Analyzer}&{QUES and UIMS}&{Change maintenance effort}&{K-fold}  \\ \hline
         
          {\cite{yenduri2021firefly}}&{FF Approach}& {RMSE}&{CKJM and IntelliJ IDEA}&{OSS Project}&{Maintainability Index}&{Leave-one-out}  \\ \hline
          
          {\cite{yenduri2021nonlinear}}&{Grey wolf Algorithm}& {RMSE}&{CKJM and IntelliJ IDEA}&{OSS Project}&{Maintainability Index}&{Leave-one-out}  \\ \hline
  \end{tabular}%
}
\end{table*}

\subsubsection{Clonal Selection Algorithm}
The clonal selection algorithm is proposed by Cutello and NicosiaIn \cite{yavuz2018prediction}. Artificial immune systems, clonal selection algorithms are a set of algorithms inspired by the clonal selection theory of acquired immunity, which explains how B and T cells enhance their affinity maturation response to antigens over time. These algorithms emphasize the Darwinian features of the theory, in which selection is motivated by the affinity of antigen-antibody interactions, reproduction by cell division, and variation by somatic hypermutation. Clonal selection techniques resemble parallel hill climbing and the GA without the recombination operator and are most frequently utilized in optimization and pattern recognition.

In 2016, Lov Kumar et al. in \cite{kumar2016hybrid} used software metrics for constructing object-oriented software maintainability prediction models. The results demonstrate the variance of MMRE in relation to antibody sizes for UIMS and QUES, respectively. Clearly, FLANN-CSA (FCSA) approach using feature extraction data using PCA in QUES and reduced attribute utilizing RST in UIMS achieved superior results. In the instance of UIMS, the model created with the entire feature set instead of the rough set yielded superior results. In the instance of QUES, the model developed utilizing the PCA feature extraction approach produced superior results. It may be concluded that FLANN-CSA requires a minimal number of generations to fulfill the stopping requirement, i.e., 95 percent of antibodies reach the same fitness value after a minimum number of generations. The findings also revealed that FCSA (RST) in UIMS provides more accurate estimations than other models used in their research.

In summary, the antibody candidate set generated by the CSA has just a small number of antibodies with high antigen affinity in order to generate mutations with a high frequency. Some low-affinity antibodies are replaced with fresh antibodies in preparation for the subsequent clonal selection. It is difficult to participate in clonal selection because a large number of antibodies with high affinity exist in antibody concentration for an extended period of time. This portion of inactive antibodies creates a "black hole" in the antibody set that is difficult to remove and update in a timely way, hence slowing the algorithm's approach to the optimal solution. This increases the complexity of the process of software maintainability prediction. The CSA also poses Challenges related to the justification of predictions of software maintainability and privacy of the data.

\subsubsection{Gene Expression Programming}
Gene Expression Programming is proposed by Ferreira in the year 2002 \cite{aslam2022compressive}. Gene expression programming (GEP) is a member of the evolutionary algorithm family and is closely connected to genetic programming and GA. It inherits the linear chromosomes of fixed length from GA and the expressive parse trees of various sizes and forms from genetic programming. GEP is an evolutionary algorithm that generates computer programs or models. These programs are complex tree structures that learn and adapt by altering their sizes, forms, and composition, similar to a live thing. Similar to live organisms, GEP computer programs are encoded in simple, fixed-length linear chromosomes. Thus, GEP is a genotype-phenotype system, with a basic genome to store and transmit genetic information and a sophisticated phenotype to explore and adapt to the environment.

Tarwani et al. in \cite{tarwani2016predicting} proposed the use of GEP for the software maintainability prediction and evaluate its performance over several ML techniques, including Decision Tree Forest, SVM, Linear regression, MLP, and Radial basis neural network. The empirical investigation is undertaken using four open-source datasets. The identification of eleven bad smells is considered a maintenance task. This study demonstrates that the GEP approach outperforms ML classifiers; hence, it is a viable alternative for predicting software maintainability.

In summary, GEP takes the advantages of both the optimization and search techniques based on genetics and natural selection, which helps in better prediction of software maintainability than GA. GEP has challenges with convergence and a lot of computation. It also has the issue of justification of predictions and data private.

\subsubsection{Firefly Algorithm}
The firefly algorithm (FA) was developed by Xin-She Yang in 2008 \cite{yang2020firefly}. The FA imitates the social behavior of summertime fireflies. Specific characteristics of fireflies include communication, prey searching, and mating. Fireflies are unisexual, fireflies are attracted to each other regardless of gender. Attractiveness is directly proportional to brightness, the lower-brightness firefly approaches the higher-brightness firefly. If no other firefly is brighter than the present one, it will go aimlessly throughout space. The relationship between brightness and the cost function is significant. In maximizing problems, the brightness is proportional to the value of the cost function.

In 2021, Gokul et al. \cite{yenduri2021firefly} proposed firefly based maintainability index by taking into account certain software metrics with the aim to minimize error. The FA is compared to other traditional models such as DE, Artificial Bee Colony, PSO, and GA in terms of performance metrics such as differential ratio, correlation coefficient, and RMSE. The proposed model has shown promising outcomes. The proposed firefly method outperformed both existing conventional and meta-heuristic models.

In summary, FA is more effective at handling multi-modal optimization problems and nonlinear problems. Additionally, it does not employ velocity. Therefore, there is no issue with velocity variation. Concurrently, FA has a rapid convergence rate for locating global optimization. In addition, similar to other GA, FA does not require the optimal initial solution to initiate the iteration phase. In addition, it may be easily combined with other optimization approaches to build hybrid algorithms. It has issues related to high computational complexity, convergence speed, justification of results and privacy of the data shared.

\section{Challenges and Future Directions.}
\begin{figure*}[h!]
    \centering
	\includegraphics[width=.8\textwidth]{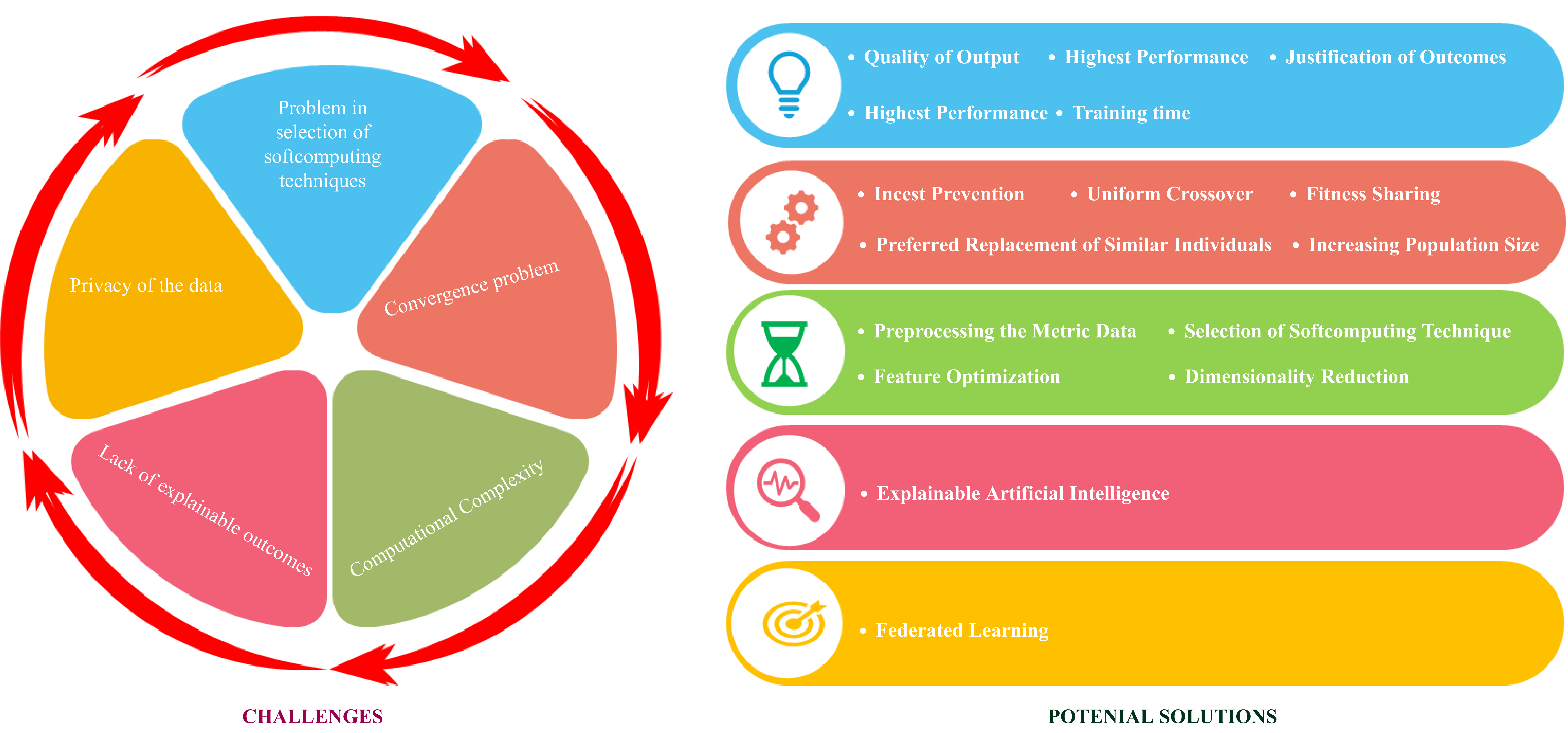}
    \caption{Challenges and Potential Solutions}
    \label{fig:CP}
\end{figure*}

The challenges, along with future directions, are summarized in fig. \ref{fig:CP}

\label{chall}

\subsection{Problem in Selection of Soft computing Techniques}

\textbf{Challenge:} The application of soft computing approaches in the prediction of software maintainability has only been the subject of a relatively small number of recognised publications. According to the findings of our investigation, GA that combine ML techniques are not only widely used but also produce superior outcomes.The comparative methods that were used in these studies are also relatively traditional in their approach. When determining the maintainability of software, other contemporary soft computing techniques, such as Harris Hawks and Honey Badger techniques, are not taken into consideration. This raises a substantial challenge on which bias the soft computing techniques are selected for prediction of software maintainability. As a result, this creates confusion among new researchers in considered soft computing techniques for software maintainability predication.\\

\textbf{Potential Solution:}
In order to select a soft computing technique for software maintainability prediction, it is crucial to consider the technique's quality of output. Priority should be given to soft computing techniques with the highest performance \cite{aziz2022application}. When selecting a soft computing technique, it is also important to consider the outcome's explanation. The complexity of the soft computing method should also be taken into account, as the higher the complexity, the more challenging it is to obtain results. The amount of data required by the soft computing technique and its processing capacity are also significant factors when adapting a soft computing technique. When choosing a soft computing technique, it is important to consider both the convergence speed and the training time.

\subsection{Convergence Problem}
\textbf{Challenge:}
It is possible for progression to come to a halt due to the convergence of soft computing in predicting software maintainability. This is because each and every individual in the population is identical. Many different soft computing techniques like GA and PSO have the potential to experience full convergence, which can result in inaccurate results. When predicting software maintainability, using soft computing techniques can lead to premature convergence, which means that the techniques have converged on a single solution, even though that solution might not be accurate. This raises a serious Challenge when considering soft computing techniques in the process of software maintainability prediction.\\

\textbf{Potential Solution:}
Incest prevention is a mating strategy that can prevent convergence \cite{altarabichi2022fast}. Additionally, the uniform crossover can prevent convergence problems \cite{ozsoydan2022hybridisation}. The preferred replacement of similar individuals through pre-selection or crowding can assist in avoiding convergence \cite{zeitrag2022surrogate}. By segmenting individuals with comparable fitness through fitness sharing, convergence can be avoided. Increasing population size will also decrease the likelihood of convergence.

\subsection{Lack of Explainable Outcomes}
\textbf{Challenge:}
In comparison to other individual models, the use of hybrid models that combine soft computing and ML techniques has yielded promising results in software maintainability predictions. However, the results of these models are ambiguous, making it difficult for humans to understand them. In order for these techniques to be trusted, the results must be justifiable and explainable, which is challenging.\\

\textbf{Potential Solution:}
Explainable artificial intelligence (XAI) is a potential solution to the justification challenge that soft computing techniques is used along AI models in predicting the maintainability of software. XAI is a collection of processes and methods that enable users to interpret and have trust in the results \cite{arrieta2020explainable,srivastava2022xai,wang2021explainable} . XAI will make software maintainability predictions generated by soft computing techniques with help of AI models more accountable. XAI can be employed to describe a outcome, its expected consequences, and any possible biases. It helps describe  correctness, fairness, and transparency which results in improved decision-making.

\subsection{High Computational Complexity}
\textbf{Challenge:}
The computational complexity of soft computing techniques is relatively high in comparison to other techniques. The soft computing algorithms must evaluate multiple sets of solutions before arriving at the global solution with the highest fitness value. The process is iterative and involves voluminous procedures and computations, necessitating vast computing and resource capacities. As a result, the selected iterations result in slow outcomes \cite{sharma2021comprehensive}. This presents a challenge in adapting soft computing techniques to the software maintainability prediction process.\\

\textbf{Potential Solution:}
Computational complexity of soft computing techniques' for software maintainability prediction can be reduced by selecting the appropriate metric data for training the soft computing techniques. The metric dataset must be pre-processed, and related to the programming paradigm in which the software is developed. The process of population initialization also plays a important role in the computational complexity, as selecting a large population may increase the complexity of the soft computing technique in predicting software maintainability. The techniques like diagonal linear uniform initialization can be used to reduce the computational cost \cite{li2021improved}.  

\subsection{Privacy of The Data}
\textbf{Challenge:}
In order to generate results, soft computing techniques will require access to a software maintainability dataset with related metrics. The user must share their data with a central server in order to generate results. There is no reliable mechanism for protecting the privacy of sensitive data shared between the user and the central server. This is the most significant challenge which creates the lack of software maintainability datasets. In order to protect privacy, there is a need for an approach in which data should not be shared but results can be processed.\\

\textbf{Potential Solution:}
Federated learning (FL) has emerged as a possible technique for facilitating distributed collaborative learning without disclosing original training data. FL enables the development of cross-enterprise, cross-data, and cross-domain applications that adhere to data protection regulations \cite{zhang2021survey,alazab2021federated}. It is a potential solution for predicting software maintainability without data sharing with the central server.

\section{Conclusion}
The goal of this literature review is to assess the significance of soft computing techniques in predicting software maintainability. In this process, we also analysed the software maintainability prediction measurements, metrics, datasets, evaluation measures, and tools. This extensive search was conducted in online digital libraries to identify journal or conference articles with peer review. The intent of this SLR is to examine all relevant research evidence and identify studies in the field of predicting software maintainability using soft computing techniques. It is understood that soft computing techniques improve the prediction of software maintainability when combined with AI techniques. In addition, it is noted that soft computing techniques face significant obstacles relating to the identification of appropriate soft computing techniques for software maintainability prediction, convergence problems, high computational cost, lack of explainability and justification of results, and data privacy concerns. This SLR also found that the use of soft computing techniques in the process of predicting software maintainability is extremely limited. This SLR suggests with help of modern soft computing techniques can assist future researchers in making more accurate predictions related to software maintainability.

\bibliographystyle{IEEEtran}
\bibliography{ref}
\end{document}